# Knowledge-Based Innovation Systems and the Model of a Triple Helix of University-Industry-Government Relations

Loet Leydesdorff •

Science & Technology Dynamics, University of Amsterdam

Amsterdam School of Communications Research (*ASCoR*),
Kloveniersburgwal 48, 1012 CX Amsterdam,
The Netherlands

Tel.: +31- 20- 525 6598; fax: +31- 20- 525 3681
loet@leydesdorff.net; http://www.leydesdorff.net/

**Abstract**

The (neo-)evolutionary model of a Triple Helix of University-Industry-Government Relations focuses on the overlay of expectations, communications, and interactions that potentially feed back on the institutional arrangements among the carrying agencies. From this perspective, the evolutionary perspective in economics can be complemented with the reflexive turn from sociology. The combination provides a richer understanding of how knowledge-based systems of innovation are shaped and reconstructed. The communicative capacities of the carrying agents become crucial to the system's further development, whereas the institutional arrangements (e.g., national systems) can be expected to remain under reconstruction. The tension of the differentiation no longer needs to be resolved, since the network configurations are reproduced by means of translations among historically changing codes. Some methodological and epistemological implications for studying innovation systems are explicated.

**Keywords:** innovation, knowledge-based systems, knowledge production, national systems, translation, Triple Helix

---

• I am grateful to Henry Etzkowitz and Terry Shinn for our ongoing discussions about the Triple Helix model.

**Introduction**

Various scholars have proposed categories for the analysis of the ongoing changes in research and innovation systems. Gibbons *et al*. (1994), for example, distinguished between 'Mode 1' and 'Mode 2' in the production of scientific knowledge. 'Mode 1' research can be considered as disciplinarily organized, while 'Mode 2' research is mainly legitimated and organized with reference to contexts of application. Rip & Van der Meulen (1996) used the concept of a 'post-modern research system' to describe S&T policy systems. National systems of innovation (Lundvall, 1988; Nelson, 1993) have been particularly studied because of their relevance for government policies (Nelson, 1982; Irvine & Martin, 1984; Freeman, 1987).

Other policy analysts have argued that systems of innovation can no longer be stabilized (like in national systems of innovation), since they remain fundamentally in transition (e.g., Cozzens *et al.,*1990). Or should one perhaps consider 'Research, Technology, & Development' (RTD) systems as in an increasingly steady state (Ziman, 1994)? Sometimes various claims are made within a single text or the contradicting statements are combined in edited volumes.

What generates this lack of consensus about *the appropriate unit of analysis* in the study of technology and innovation? In my opinion, the various metaphors in the study of knowledge-based systems (Wouters *et al.*, 1999; Cutcliffe, 2000) can be considered as theoretical appreciations of a complex dynamics from different perspectives and with potentially different objectives. The analysts attempt to stabilize a picture by choosing a perspective. Since the systems under study are developing dynamically, the metaphors remain 'out of focus' when viewed from one window of appreciation or another. However, the illusion of a stable object enables policy advisors to legitimate S&T policies.

Given the complexity of the dynamics unintended consequences (e.g., economic externalities) can be expected to prevail (Giddens, 1984; Callon, 1998). Innovation is not a stable unit of analysis, but a unit of operation at an interface. The various story lines show the symmetry breach at the interface. One is able to provide reflections on the operation at an interface from different angles or at different moments in time.

From this meta-perspective, knowledge-based developments can be considered as systems that are (discursively) reconstructed while (recursively) developing *in interaction with* other subsystems of society, e.g., markets. The dynamics of such non-linear interactions are non-trivial and nearly incommensurable windows of appreciation can be expected. In this contribution, I foreground the self-organizing potentials of the complex dynamics of innovation and the function of the variety of reflections on this phenomenon (Leydesdorff, 1997).

**The complex dynamics of innovation**

While each innovation can be considered as an instance of interaction, innovation *systems* build recursively on the interaction terms. Because these systems are constructed on the basis of interactions at interfaces, they can be expected to exhibit non-linear dynamics. Furthermore, the recursion in the interaction terms remains beyond control when analyzed from the perspective of either of the subsystems that interact. In other words, a model has to contain both interactive and recursive terms at different levels of aggregation.

In such a non-linear model, intentional input can no longer be expected to lead to intended output. The dependency relations themselves can be expected to change when the systems under study are further developing. 'Externalities' emerge in unexpected contexts. Contexts become crucial when consequences can no longer unambiguously be related to causes (e.g., Barnes & Edge, 1982). While policy makers may be able to steer the developments in some cases and at some stages, the policy inputs can be expected to remain contextual in other instances—for example, when the systems internalize complexity by becoming increasingly knowledge-intensive (Van den Daele *et al.*, 1979).

The externalities of innovation systems can be made visible only reflexively, that is, *ex post*. From this perspective of hindsight, the focus of the analysis shifts from the social construction of technology (e.g., Bijker *et al.*, 1987) to the complex dynamics of innovation in the constructed system. The social construction by agency can then be considered as the subdynamics of variation, whereas selection is structured at other levels. Each subdynamics—that is, specific interactions—can be made the subject of theoretical reflections by taking a perspective or making specific assumptions; for example, about the role of agency *or* structure (cf. Giddens, 1984).

In general, the theoretical discourses attempt to stabilize a geometrical representation of the algorithmic dynamics of interactions and fluxes under study. In addition to the theoretical task of improving on the quality of the theoretical reflections, the methodologist is then able to raise questions concerning the contingent relations between different forms of appreciative theorizing. What is the relative quality of the different reflections? Is one able to appreciate also the surplus value of the interaction among the various perspectives?

Note that innovation itself can be considered as a reflexive recombination from different perspectives. However, innovation is then defined at the level of the systems under study. The perspective of providing feedback by 'recombination' at the reflexive level relates the study of knowledge-based innovation systems with the complexity approach. The model of evolutionary theorizing in economics can then be recognized as providing a meta-biological perspective in which selection environments, for example, have often been considered as 'given'.

Evolutionary economists have first drawn attention to the non-linear interaction terms between market perspectives and technological options (Mowery & Rosenberg, 1979;

Freeman, 1982). From a sociological perspective, however, neither market perspectives nor technological options are biologically given like the 'genotypes' of an organism. All interacting subsystems at the social level (technologies, markets, institutions, etc.) have discursively been constructed and are continuously reconstructed.

For example, Nelson (1994) proposed to analyze co-evolutions between technologies and institutions in addition to co-evolutions between markets and technologies. However, institutional selection operates very differently from the dependency relations between technologies and markets (McKelvey, 1996). The Triple Helix thesis focuses on the interactions among these various interfaces. How are organizational rigidities among them organized and dissolved? When can these reorganziations be considered as structural adjustments to technological developments (Freeman & Perez, 1988)? How does a system of innovations build on stability *and* change? In which phases can change and/or stability be expected to prevail, and why?

The geometrical perspectives on these complex dynamics can be updated by unexpected interaction effects in the systems under study. These systems, however, cannot be defined without theorizing. They are not given, but historically constructed. Therefore, they can be further refined by reflections that are stabilized at the social level. The codification of these reflections is stimulated when the respective interpretations are disturbed by interactions. Thus, the various metaphors function both as evaluation schemes and as heuristics. The innovation 'system', however, is knowledge-based by definition.

While studying innovation systems one becomes increasingly aware of the dependency of each analytic perspective on its definitions. Changes in definitions sometimes provide windows of unexpected opportunities for innovation. Different stakeholders (e.g., academia, industry, government) recombine from their respective perspectives. Recombinations (that is, knowledge-based innovations and reorganizations) potentially disturb the current discourses to such an extent that new perspectives can be proposed and sometimes elaborated.

**Expectations, institutions, and communications**

The institutional arrangements under study (e.g., national systems of innovation) compete in terms of their respective successes and failures when attempting to grasp the fruits of possible innovations, for example, by trading off changes in their structures (transaction costs) against historical continuities (routines). Thus, the various subdynamics operate upon one another without any *a priori* guarantee of harmonization, but under selection pressure. Order can be expected to emerge in one or more directions because of potential resonances between the selecting systems (Kampmann *et al.*, 1994). These 'lock-ins' (Arthur, 1988, 1989, and 1994) remain conditioned and constrained by the historical configurations (David, 1985).

When and where do the emerging conditions fit into each other, and to what extent? The metaphor of an overlay of mutual expectations and exchange relations enables us to analyze these complex dynamics as a result of the interaction among the various subdynamics, while each of the subdynamics can also be declared as recursively operating on the basis of their previous state. Thus, the differentiation is also reproduced. Furthermore, the analyst may act as a participant (recursively implied) and/or as an external observer. One is additionally able to change one's perspective, for example, when giving normative advice.

Corresponding to this double perspective of analyst and participant (Giddens, 1976), the emerging overlay can be considered by each actor as subsystemic (that is, as an interface *within* the system) and/or as supersystemic (that is, as a factor in the system's environment). While a supersystemic factor provides a relevant environment for the system of reference, each participant can also be implied in the (re)construction of the overlay by reflecting on his or her environment. Thus, a double perspective of participant and observer is reflexively reinforced and knowledge-based learning processes are then induced (Leydesdorff, 2001a).

From this (neo-)evolutionary perspective, the observable social structures can be considered as successful cases of previous institutionalization and conflict resolution. The structural forces behind the institutionalization and stabilization remain latent, but they can be hypothesized. In the longer term, institutions can be expected to optimize their relations with relevant environments (for example, by learning to cope with uncertainties). Thus, the knowledge base of these institutions is further developed. Institutional functionality in a knowledge-based economy, however, also implies reaching across institutional borders on the basis of expectations about how the environments can be expected to change.

For example, industries have to assess in what way and to what extent they decide to internalize R&D functions. Universities position themselves in markets, both regionally and globally. Governments make informed trade-offs between investments in industrial policies, S&T policies, and/or delicate and balanced interventions at the structural level. Such policies can be expected to be successful insofar as one can anticipate and/or follow trends according to the dynamics of the new technologies in their different phases (Freeman & Perez, 1988; McKelvey, 1996; Giesecke, 2000). The management of these interfaces is both an economic imperative and a political challenge, yet knowledge-intensive in the elaboration.

Is this (neo-)evolutionary model a reappraisal of old-fashioned structural-functionalism? In my opinion, the Triple Helix model extends the basis of structural-functionalism by introducing the notion of 'meaning' from symbolic interactionism: social functions are discursively constructed, and they can be deconstructed and reconstructed reflexively. Thus, one can no longer accept a dialectics between ahistorical functions and historical institutions. The institutions are needed to carry out the functions, but they can be expected to be changed while doing so. The functions are continuously under reconstruction and the institutional elements of the systems have been generated by these reflexive operations.

**The historical (re-)construction**

The evolutionary model is historically reflexive, since the (cultural) evolution builds on the achievements of the past. Both layers of the system (institutions and functions) have been socially constructed and stabilized, but in different periods of time. First, the communications were functionally differentiated as in the individual revolutions of the 16$^{th}$ and 17$^{th}$ century. The transformations of the 18$^{th}$ century ('modernization') led in the 19$^{th}$ century to the institutional differentiation between the modern state and civic society.

The complex social system builds on the interfaces among institutions and functions as different mechanisms of differentiation. After the completion of a system of nations, that is, from approximately 1870 onwards (Noble, 1977), the interactions between institutions and functions could gradually be reconstructed into a knowledge-based system. Because of the increasing knowledge-intensity of the communications, one is increasingly able to experiment with the interaction terms between structures and functions in the organization of social systems.

Institutions can be assessed in terms of their functionality, and functions can be evaluated in terms of their value for the carrying institutions. Functional meanings and institutional meanings can constructively be 'translated' into each other. Since the interaction terms are based on reflections and therefore not always readily observable, they suppose the declaration of reflexivity as a condition for their discursive reconstruction. This discursive reflexivity can always be made more knowledge-intensive and science-based.

Let me follow the sociological tradition (Marx, Weber, Parsons) in assuming that the functional differentiation of society was constructed during the individual revolutions of the 16$^{th}$ and 17$^{th}$ centuries. Only after the completion of the 'modern' system (that is, 'nature' as the subject of mathematical physics) during the scientific revolution of the late 17$^{th}$ century could institutional implementation of this 'natural' and 'universal' system be legitimated as deliberate reform of social relations ('modernization'). Foucault (e.g., 1972) used the term 'noso-politics' for this reconstruction of society during the Enlightenment of the 18$^{th}$ century.

The institutional differentiation of the state from civic society followed upon the American and French Revolutions and was then achieved in the first half of the 19$^{th}$ century. This development has led to a variety of nation states with their respective cultures. From 1870 onwards, a scientific-technological revolution can subsequently be distinguished, gradually shaping a knowledge-based mode of production and distribution at the global level (Braverman, 1974).

During this latter process the knowledge production and control functions increasingly ceased to be the exclusive domain of academia (Noble, 1977). Functions and institutions can historically be coupled, but there are no determinate relationships (Whitley 1984). For example, the American university, to a larger extent than its German predecessor, became also an entrepreneurial locus (Etzkowitz, 2001). The position of government changed from that of a principal agent ('King') into that of a controlled bureaucracy negotiating an internal trade-off between facilitating further developments at the level of society and political accountability (e.g., Van den Belt & Rip 1987; cf. Weber 1922).

| *Functions* *Institutions* | *Science* | *Economy* |
|---|---|---|
| *Public* | **Academia; University** | Patent legislation; Science, technology, and innovation policies |
| *Private* | Industrial R&D labs; entrepreneurial universities | **Trade and Industry** |

**Table 1**
*The interaction between functional and institutional differentiation*
*(since approximately 1870)*

In the liberal organization of society, science was first considered as a public good, while trade and industrial production were considered private activities. These categories became increasingly interchangeable across institutional interfaces with the further development of the system. Scientific insights could be made useful in industrial practices and industrial (and military) concerns began to guide the heuristics of scientific research programs. These reflexive flexibilities, perhaps rooted in American pragmatism, also influenced the construction of the European Union after W.W. II, since a variety of perspectives must be translated into each other in order to (re-)construct the European dimension (cf. Ronge, 1979).

**The Triple Helix as a post-institutional model of cultural evolution**

The new mode of knowledge-based production can be expected to build on the old one(s) as its historical basis. Thus, labels like 'university', 'industry,' and 'government' did not disappear while constructing the transnational system, but they gradually shifted in meaning (Callon & Latour, 1981). However, the changes of meaning are not expected to imply a loss of differentiation at the level of the systems under (re)construction. The differentiation hitherto obtained by the social system provides its complex dynamics with the capacity to develop further in response to emerging challenges.

From this perspective, the established systems and their corresponding discourses compete as suboptimal solutions to the problem of organizing and giving meaning to a social world (a 'universe'). Nation systems, for example, can be considered as cultures which compete for a share in the economic developments at the global level. One no longer expects a single (optimum) solution based on an undifferentiated integration, a common center or a textbook (as in a high culture; cf. Yamauchi, 1986). One expects local suboptima which explore 'hill-climbing' in their relevant environments (Kauffman, 1993; Allen, 1994).

Since social systems remain distributed by their nature, the institutional hill-climbers compete and thereby reshape the distribution of their opportunity structures in relation to one another. Thanks to this reshaping, the landscape itself can be expected to change (Scharnhorst, 1998). As in biology, the landscape is rugged with its historical formations. Unlike biology, however, Schumpeter's (1939) 'creative destruction' of existing constructions is part of the reflexive practices of the hill-climbers (although changing the definitions may require a lot of cultural energy). 'Nature' and 'culture' could thus increasingly be understood as dialectical categories (e.g., Marx).

In contrast to a double helix, that is, a coevolution between two subdynamics (e.g., production forces and production relations), a Triple Helix cannot be expected to be stabilized or resolved. A model of three helices is sufficiently complex for understanding the complex dynamics of the ongoing transformation processes. The (three) double helices on which the Triple Helix builds, continuously 'lock-in' into local coevolutions that are expected to 'clot' into provisional solutions shaping the ruggedness of the corresponding landscapes.

The clots of different sizes perform their own 'life'-cycles (along historical 'trajectories'), for example, at the industry level, while the landscape can be considered as a next-order system forming a 'regime' (Dosi, 1982). Bifurcations endogenously reshape the levels in series of events. The 'regime' is a non-linear effect of the trajectories (Scharnhorst, 1998), and the 'trajectories' are the expected consequences of previous lock-ins (Leydesdorff & Van den Besselaar, 1998).

In biology, the *rugged fitness landscapes* are 'given' for the various species and provide niches for their survival. In an economy, the niches can be considered as mechanisms for the retention of adapted environments: how is one able to organize an institutional arrangement so that wealth and jobs can be generated? These social mechanisms, however, can also be deconstructed and reconstructed.

The solution of local conflicts has hitherto been a central function of the nation states and their political economies (Skonikoff, 1993). As the operation of the third helix (knowledge) becomes more pronounced in the re-organization of society, it continuously may destabilize a provisional click between two other helices. Nelson & Winter (1982), for example, noted that technological innovations tend to upset the equilibrating dynamics of the market. The market, however, can only be equilibrated within an institutional setting. Destabilization can also be considered as an effect of interactions at the regime level (Freeman & Perez, 1988).

Technologies sometimes 'click' with state apparatuses into a local fit, like in the energy sector or in health care; industries potentially click with technologies (e.g., David's (1985) QWERTY keyboards and Arthur's (1988) VHS tapes); and industries can click with the state apparatus as in the former Soviet Union. A click excludes a third subdynamic from effectively operating, since the co-evolutionary dynamic can then be considered as temporarily 'locked' (Simon 1969).

For example, the political economies of Eastern Europe were not sufficiently able to make the transition to a knowledge-based economy during the 1970s and 1980s (e.g., Richta *et al.*, 1968). When the Chinese innovation system was confronted with similar problems of integration in the late 1980s, one could reflexively choose for a knowledge-based reform of the political economy (Leydesdorff & Guoping, 2001).

A lock-in can have local advantages (e.g., increasing marginal returns) and/or global disadvantages. A 'break out' of a lock-in may open a window on a new market and offer a global (that is, next-order) perspective. Sometimes it provides also a niche for developing a new discipline. ICT (e.g., Nowak & Grantham, 2000) and biotechnology (e.g., McKelvey, 1996) have been considered as prime examples. But the risk of crisis is ever present given the complexity of the dynamics. As another dynamics increasingly disturbs a local harmonization, the systems can be expected to become 'critical', that is, to drift towards the edges of chaos and bifurcation.

**The predictive power of the neo-evolutionary model**

Can a reflexive observer grasp the evolutionary momentum and perform the adjustments in time or not? What unintended consequences can be made visible *ex ante* using the available reflections? How large are the expected uncertainties? Can the threats also be formulated in terms of opportunities?

As noted, the differentiations that have been achieved historically, cannot be dissolved at the system's level without costs. A loss of internal complexity, for example, can be expected to lead to a loss of ability to handle complexity in the relevant environments (e.g., markets). The functional differentiations of knowledge production, wealth creation, and governance operate as feedbacks on institutional task divisions among academia, industry, and government, and *vice versa*. However, one can no longer expect a single or pre-given (e.g., national) order to prevail: the various subdynamics are juxtaposed. Whether, when, and where they lock-in, remain historical questions. Thus, there is always room for improvement, empirical investigation, and change.

When order can be observed, the analyst may be able to indicate on theoretical grounds how this order was constructed, for example, at the level of nations or sectors. Simulation studies enable us to specify the (sometimes counter-intuitive) expectations given historical configurations and theoretical assumptions about the relevant subdynamics. However, the simulation results require another round of appreciation.

Although the analyst may be able to specify the uncertainty, the assumption of the non-trivial (social) machinery of a Triple Helix with an overlay *adds* to the uncertainty. For example, other players may see options for codification that the analysts could not possibly have seen given their necessarily contingent positions as also participants. Reflexivity and uncertainty prevail in a knowledge-based economy.

Biological evolution theory assumed variation as the driving force and selection to be 'naturally' given. Cultural evolution, however, can be considered as driven and reconstructed both by individuals and groups who make conscious decisions and by the appearance of a variety of unintended consequences of interactions with which one may have to cope in a next stage. Since the sources of innovation in a Triple Helix configuration are not synchronized *a priori*, the possibilities for innovations and rearrangements generate puzzles for participants, analysts, and policy-makers alike.

This network of reflexive relations operates as a knowledge-intensive subdynamics of intentions, strategies, and projects that adds surplus value by reorganizing and harmonizing the political and economic structures in order to achieve a better approximation of the variety of (uncertain) goals. The issue of how much and under which conditions anyone is in control given this layer of interacting expectations specifies a research program for innovation studies.

In the case of innovation systems, however, the expectation is that 'what you see is *not* what you get'! What you see, are the footprints of previous operations. The definition and consequently the delineation of systems is knowledge-intensive. The interacting subdynamics, that is, specific operations like markets and technological innovations, are continuously reconstructed—like e-commerce on the Internet—yet differently at different places and various levels. What is considered as 'industry' and what as 'markets' cannot be taken for granted and should not be reified. Each 'system' is defined and can be redefined as research projects are further designed.

For example, 'national systems of innovation' can be more or less systemic. The extent of systemness can be studied as an empirical question (Leydesdorff & Oomes, 1999; Leydesdorff, 2000). Dynamic 'system(s) of innovation' may consist of increasingly complex collaborations across national borders and among researchers and users of research from various institutional spheres (Godin & Gingras, 2000). Among other things, one may expect differences among regions (Braczyk *et al.*, 1998) and sectors (Pavitt, 1984).

All these systems of reference can be specified analytically, but their systemness remains a hypothesis. The Triple Helix hypothesis states that the 'systems' can be expected to remain in transition (Etzkowitz & Leydesdorff, 1998). Can the observations then still provide an opportunity to update the expectations?

**The status of the observables**

Non-linear models of innovation extend linear models by taking interactive and recursive terms into account. The non-linear terms can be expected to change the causal relations between input and output, that is, the production rules in the systems under study. The reflexivity in the discourses adds a layer of learning to these evolving systems. These distributed systems (at the network level!) can be both hyper-reflexive (Woolgar, 1988) and infra-reflexive (Latour, 1988) when operating (cf. Ashmore, 1989).

The reflections remain structurally coupled to the actors involved as the carriers, but one sub-dynamics may be repressed (deselected) at the social level more than another given historical contingencies. By changing the unit of analysis or the unit of operation reflexively, one is able to obtain a different perspective on the systems under study. But the latter are evolving at the same time. In terms of methodologies, this continuous change at various levels challenges our conceptual apparatus, since one is not always able to distinguish clearly whether a variable has changed ($dx/dt$) or merely the relative value of this variable ($x$).

Discursive analyses provide us with snapshots, while reality presents a moving picture. The analysts, however, need geometrical metaphors to render the complexity accessible to understanding, and these metaphors can be stabilized by higher-order codifications, as in the case of paradigms. An understanding in terms of fluxes (that is, how the variables as well as the values change over time) calls for the use of algorithmic simulations. The observables can then be considered as special cases which inform the expectations over time (Leydesdorff, 1995).

The study of innovation systems requires this level of sophistication, since innovations can be defined only in terms of an operation which one can expect to contain both recursivity (stability) and interactivity (change). Knowledge-based innovation is therefore a cultural achievement: the innovators themselves are reflexive with respect to previous solutions. Both the innovator(s) and the innovated system(s) are expected to be changed by innovation. As noted, one is able to be both a participant and an observer, and one needs to be able to change these perspectives reflexively in order to maintain a position within this process.

Although the different narratives are mixed and confused in 'real life' events, the various models can be distinguished analytically. In his study of 'artificial life', Langton (1989) proposed to distinguish between a 'phenotypical' level of the observables and a 'genotypical' level of analytical theorizing. The 'phenotypes' remain to be explained, while the various theories compete in terms of their 'idealtypical' clarity and their usefulness in updating the respective expectations. Confusion, however, is difficult to avoid given the 'real life' pressures to jump to conclusions. The different perspectives are continuously competing, both normatively and analytically.

**Epistemological implications**

The innovation systems under study can be expected to contain a complex dynamics and therefore, they do not have to be integrated nor completely differentiated. On the contrary, they are further developing. Under selection pressure, however, they can be expected to perform on the edges of fractional differentiations and local integrations.

Using this model of partial and temporary solutions, one is able to understand why the knowledge-based regime exhibits itself in terms of progressive instances and non-periodical crises.

The local sequences inform us about global developments in terms of the deselected exceptions which can be replicated and built upon. The selection mechanisms, however, remain theoretical constructs. Historical case studies provide us with positive instances that enable us to specify these (negative) selection mechanisms, but only reflexively. Over time, the initial inferences can perhaps be corroborated. At this end, the function of reflexive inferencing based on available and new theories moves the system forward by drawing attention to possibilities for change.

While integration works in a hierarchical mode, reflexive translations can tolerate inconsistencies and differences to be resolved over time. The ensuing puzzles set a research agenda. In general, translations operate among heterarchical (since different) codifications. The exchanges at the interfaces enable us to transfer insights from one social domain to another by placing them in a different context, and thus by providing the substantive information with another meaning.

In a functionally differentiated system, each message is expected to contain an information which can be provided with specific meaning according to the code which is functional for the development of this (sub)system (Luhmann, 1984). For example, in science one investigates whether a given statement is true or false, while in the economy one assesses whether one can utilize a finding (e.g., the patent) to make a profit. For the latter purpose, one does not always have to understand the underlying mechanism in great detail. However, one cannot buy the 'truth' of a statement on the market or claim it with only political power (Lecourt, 1986).

When institutional differentiation (in the retention mechanism) is added to the functional differentiation in the exchange, the theoretical specification becomes one step more complex. As noted, this expansion of the economy became urgent from 1870 onwards, when industrial R&D laboratories were installed and when patent legislation increasingly allowed for the transfer of insights from the laboratory into practice and *vice versa*. In principle, the two operations—of functional differentiation and institutional transfer—can operate as selectors upon each other, but only at certain places can resonances into coevolutions ('mutual shapings') be expected. Thus, in most places the options will not match, but, in a scattered and distributed mode, a new mode of knowledge production can be generated. The local resonances compete at a next-order systems level. In order to study this selection, this next-order system has then fist to be hypothesized and specified.

In the case of an emerging network, each communication can increasingly have a meaning for systems other than the one in which it was generated. In addition to their intrinsic meaning, the communications can then be valued systematically from different, but related perspectives. Since communications can be communicated, these networked systems generate internal complexities which require interface management. Galbraith's (1967) 'techno-structure' can be considered as a first solution to this problem. What competencies belong to which domains? As the organization of the interfaces and the control functions within them then also become degrees of freedom, the unambiguous attribution of a communication to either of the systems becomes increasingly difficult.

A comparison may be helpful here: translations of texts into other languages can be unambiguous because the grammars and dictionaries of the various languages are codified. Thus, if one feels uncertain, one can look up the correct translation in a dictionary or one may ask a native speaker. The situation among functional domains, however, is one in which the translators are discussing about the proper translations among themselves. When the translators happen to agree, a new codification may be more useful than the 'natural' (that is, previously given) ones.

In a complex dynamics, each system remains under reconstruction and in evolutionary competition while reorganizing complexity within its relevant environments. Since all these systems are increasingly knowledge intensive, their borders remain uncertain expectations. New codifications reconstructed by the translations may then become more functional than the originally translated ones (or not).

The philosophy of science has been responsive to these developments. First, the systematic use of science in industry in the late 19$^{th}$ century raised fundamental questions about the *demarcation* between science and non-science at the interfaces. This issue was solved by the so-called 'linguistic turn in the philosophy of science' during the interbellum. While previously truth had been associated with ideas, a truth-value would henceforth be attributed to statements: some statements are more likely to be true than others.

The post-modern turn has changed the situation again: the truth-value of a statement is also contextual. One has a degree of freedom to play with the centrality of concepts in terms of heuristics and puzzle solving (Simon, 1969). As Kuhn (1962) noted, the precise definition of 'atomic weight' may differ between chemical physics and physical chemistry without necessarily creating confusion. Concepts have meaning within discourses. Translations between discourses and reformulations within discourses can nowadays be considered as the carriers of knowledge-based developments.

This implies neither arbitrariness in what is true or not, nor a relativistic position. The various values of a communication (including its potential truth) can only be discussed from within a discourse. These discourses are themselves developing and thus changing in terms of what is true or not. Although the discourse is uncertain in terms of its boundaries, it can also be expected to be more certain in terms of its core. Codifications structure the discourses, and translations enable us to communicate among them.

Note that the translations are selective and asymmetrical. One cannot force validity upon a scientific discourse from a political perspective. Analogously, the validity of a scientific statement does not guarantee its diffusion. The systems of reference have to be specified first, and then one can raise questions with respect to quality control in these various dimensions (that is, within and among these systems).

For example, one can distinguish between (*i*) validation problems which are generated within the scientific communication system because of the differences between 'Mode 1' and 'Mode 2' research, and (*ii*) validation problems within the knowledge produced in applicational contexts (Fujigaki & Leydesdorff, 2000). New mechanisms of quality control can sometimes be expected to emerge (for example, at the Internet). Thus, the new mode of the production of scientific knowledge can empirically be operationalized in an epistemological domain, that is, with reference to the validity of scientific knowledge, yet without harming the standard conventions of scientific soundness in terms of true and false statements (Leydesdorff, 2001b).

**Methodological consequences**

Since the (reflexive) selection mechanisms of cultural evolution cannot be identified by unmediated observation—note that my position can therefore be considered as anti-positivist—the neo-evolutionary analysis has to begin with the specification of a hypothesis. 'What' does one expect to be communicated and why? The study of how this communication is institutionally arranged (and therefore measurable, in principle) is then a question of empirical design.

The observable arrangements thus have an epistemological status beyond merely providing the analyst with one or another, as yet unreflexive starting point for the narrative. The data can be used for informing *ex ante*—and sometimes testing *ex post*—the theoretical expectations. Which layer operated with which function, why and in which instances?

By raising first the substantive question of 'what is communicated?'—e.g., economic expectations (in terms of profit and growth), theoretical expectations or assessment of what can be realized given institutional and geographic constraints—the focus is firmly set on the specification of the media of communication. How are these media of exchange related and converted into one another? Why are these processes mutually attractive, and under what conditions can the exchanges among them be sustained?

The helices communicate recursively over time in terms of their own respective codes. Reflexively, they can sometimes take the role of each other, yet only to a certain extent. Although the discourses are able to interact at the interfaces, the frequency of such external interactions is (at least initially) lower than the frequency of interactions within each helix (Simon, 1969).

Over time and with the availability of ICT, this relation may begin to change. Codified media provide the systems with opportunities to change the meanings of communications (given another context) while maintaining their substance (Cowan & Foray, 1997). One can expect that the balance between spatial and virtual relations is contingent upon the availability of the exchange media and their respective codifications. The maintenance of codification may be costly, but it enables us to suppress noise in the communication.

Inter-human communication remains failure prone. Quality control of communication is crucial for developing the knowledge base of the system, but it remains a 'counter-factual' expectation. One can only observe the selective operation with hindsight, that is, on the basis of a theoretical inference. Despite this 'virtuality' of the structural operation at the level of the overlay (Giddens, 1984), the socio-economic system—that is potentially innovated by the operation—is not 'on the fly': it remains grounded in a culture which has to be reproduced in terms of renewing the systems of coordination. However, the retention mechanism of the social system is no longer given; the institutonal layer is increasingly 'on the move.' It can be reconstructed as the system is deconstructed, that is, as one of its subdynamics.

**Conclusion**

The emerging system rests like a hyper-network on the networks on which it builds (such as the disciplines, the industries, and the national governments), but the knowledge-economy transforms 'the ship while a storm is raging on the open sea' (Neurath *et al.*, 1929). As the technological culture provides options for recombination, the boundaries of the carrying communities can also be reconstituted. The price of changing the communal base may also be felt as a loss of traditional identities, or alienation, or as a concern with the sustainability of the reconstruction. However, 'creative destruction' entails the option of increasing development. In this sense, knowledge-based innovation can be turned into 'a celebration of community' (Hayward, 1998).

I have argued that the evolutionary perspective in economics can be complemented with a turn towards reflexivity in sociology in order to obtain a richer understanding of how the overlay of communications in university-industry-government relations reshapes the systems of innovations that are currently subjects of debate, policy-making, and scientific study. Without this reflexive turn, evolutionary economists tend to reify systems like 'national systems of innovation,' 'regional innovation systems,' or sectors of the economy. Systems, however, can always be redefined by both participants and analysts. Furthermore, the participants and the analysts use their system's definition in studying and changing the system.

Whereas a coevolution between, for example, markets and technologies can be entrained along a trajectory, a triple helix of university-industry-government relations can no longer be expected to stabilize. The global regime can be expected to contain competing trajectories. However, the analyst can hypothesize whether a technological development (e.g., aircraft) is in its trajectory phase or in the regime phase (Frenken & Leydesdorff, 2000). In the regime phase, knowledge-based innovation policies should be aimed at influencing the political economy of a technology and not merely at addressing the trajectory of its further improvements.

**References**

Allen, Peter M. (1994) "Evolutionary Complex Systems: Models of Technology Change", in Leydesdorff & Van den Besselaar (eds.), pp. 1-17.
Arthur, W. Brian (1989) "Competing Technologies, Increasing Returns, and Lock-In by Historical Events", *Economic Journal* 99: 116-31.
Arthur, W. Brian (1988) "Competing technologies", in Dosi *et al.* (eds.), pp. 590-607.
Arthur, W. Brian (1994) *Increasing Returns and Path Dependence in the Economy.* Ann Arbor: University of Michigan Press.
Ashmore, Malcolm (1989) *The Reflexivity Thesis: Wrighting Sociology of Scientific Knowledge.* Chicago and London: Chicago University Press.
Barnes, Barry, and David Edge (1982), *Science in Context.* Cambridge, MA: MIT Press.
Bijker, Wiebe, Thomas P. Hughes, and Trevor Pinch (eds.) (1987) *The Social Construction of Technological Systems.* Cambridge, MA.: MIT Press.
Braczyk, Hans-Joachim, Philip Cooke, and Martin Heidenreich (eds.) (1998) *Regional Innovation Systems.* London/ Bristol PA: University College of London Press.
Braverman, Harry (1974) *Labor and Monopoly Capital.* New York/London: Monthly Review Press.
Callon, Michel (1998) "An essay on framing and overflowing: economic externalities revisited by sociology", in Michel Callon (ed.) *The Laws of the Market,* pp. 244-269. London: Macmillan.
Callon, Michel, and Bruno Latour (1981) "Unscrewing the big Leviathan: how actors macro-structure reality and how sociologists help them to do so", in Karin D. Knorr-Cetina and Aaron V. Cicourel (eds.), *Advances in Social Theory and Methodology: Toward an Integration of Micro- and Macro-Sociologies,* pp. 277-303. London: Routledge & Kegan Paul.
Cowan, Robin. and Dominique Foray (1997) "The Economics of Codification and the Diffusion of Knowledge", *Industrial and Corporate Change* 6: 595-622.
Cozzens, Susan, Peter Healey, Arie Rip, and John Ziman (eds.) (1990) *The Research System in Transition.* Boston, etc.: Kluwer.
Cutcliffe, Stephen H. (2000) *Ideas, Machines, and Values: An introduction to science, technology, and society studies.* Lanham, etc.: Rowman & Littlefield.
David, Paul A. (1985) "Clio and the Economics of QWERTY", *American Economic Review* 75: 332-37.
Dosi, Giovanni (1982) "Technological Paradigms and Technological Trajectories: A Suggested Interpretation of the Determinants and Directions of Technical Change", *Research Policy* 11: 147-162.
Dosi, Giovanni, Chistopher Freeman, Richard Nelson, Gerald Silverberg, and Luc Soete (eds.) (1988) *Technical Change and Economic Theory.* London: Pinter.
Etzkowitz, Henry (2001) *The Second Academic Revolution: MIT and the Rise of Entrepreneurial Science.* London: Gordon Breach, forthcoming.
Etzkowitz, Henry, and Loet Leydesdorff (1998) "The Endless Transition: A "Triple Helix" of University-Industry-Government Relations", *Minerva* 36: 203-208.
Foucault, Michel ([1969] 1972) *The Archaeology of Knowledge.* New York: Pantheon.
Freeman, Christopher, and Carlota Perez (1988) "Structural crises of adjustment, business cycles and investment behaviour", in Dosi *et al.* (eds.), pp. 38-66.
Freeman, Christopher (1982) *The Economics of Industrial Innovation.* Harmondsworth: Penguin.
Frenken, Koen, and Loet Leydesdorff (2000) "Scaling Trajectories in Civil Aircraft (1913-1997)", *Research Policy* 29(3): 331-348.

Fujigaki, Yuko, and Loet Leydesdorff (2000) "Quality Control and Validation Boundaries in a Triple Helix of University-Industry-Government Relations: 'Mode 2' and the Future of University Research," *Social Science Information* 29(4): 635-655.

Galbraith, John K. (1967) *The New Industrial State.* Harmondsworth: Penguin.

Gibbons, Michael, Camille Limoges, Helga Nowotny, Simon Schwartzman, Peter Scott, and Martin Trow (1994) *The New Production of Knowledge: The dynamics of science and research in contemporary societies.* London: Sage.

Giddens, Anthony (1976) *New Rules of Sociological Method.* London: Hutchinson.

Giddens, Anthony (1984) *The Constitution of Society.* Cambridge: Polity Press.

Giesecke, Susanne (2000) "The Contrasting Roles of Government in the Development of the Biotechnology Industries In the U.S. and Germany", *Research Policy* 29(2): 205-223.

Godin, Benoît, and Yves Gingras (2000) "The Place of Universities in the System of Knowledge Production", *Research Policy* 29(2): 273-278.

Hayward, James (1998), *personal communication,* at the Second Triple Helix Conference, New York, January 1998.

Irvine, John and Ben R. Martin (1984) *Foresight in Science: Picking the Winners.* London: Pinter.

Kampmann, Christian, Christian Haxholdt, Erik Mosekilde, and John D. Sterman (1994) "Entrainment in a Disaggregated Long-Wave Model", in Leydesdorff and Van den Besselaar (eds.), pp. 109-124.

Kauffman, Stuart A. (1993) *The Origins of Order: Self-Organization and Selection in Evolution.* Oxford: Oxford University Press.

Kauffman, Stuart A. (1995). *At Home in the Universe.* Oxford: Oxford University Press.

Kuhn, Thomas S. (1962). *The Structure of Scientific Revolutions.* Chicago: University of Chicago Press.

Langton, Christopher G. (ed.) (1989) *Artificial Life.* Redwood City, CA: Addison Wesley.

Latour, Bruno (1988) "The Politics of Explanation: An Alternative", in Steve Woolgar and Malcolm Ashmore (eds.), *Knowledge and Reflexivity: New Frontiers in the Sociology of Knowledge*, pp. 155-77. London: Sage.

Lecourt, Dominique, Lyssenko (1976) *Histoire réelle d"une science prolétarienne.* Paris: Maspero.

Leydesdorff, Loet (1995). *The Challenge of Scientometrics: the development, measurement, and self-organization of scientific communications.* Leiden: DSWO Press, Leiden University; at <http://www.upublish.com/books/leydesdorff-sci.htm>.

Leydesdorff, Loet (1997) "The Non-linear Dynamics of Sociological Reflections", *International Sociology* 12(1): 25-45.

Leydesdorff, Loet (2000) "Are EU Networks Anticipatory Systems? An empirical and analytical approach", in Daniel M. Dubois (ed.), *Computing Anticipatory Systems -- CASYS"99,* pp. 171-181. Woodbury, NY: American Physics Institute.

Leydesdorff, Loet (2001a) *A Sociological Theory of Communication: The Self-Organization of the Knowledge-Based Society*. Parkland, FL: Universal Publishers; at <http://www.upublish.com/books/leydesdorff.htm>

Leydesdorff, Loet (2001b) "Indicators of Innovation in a Knowledge-Based Economy," *Cybermetrics* 5(1) Paper 2, at
< http://www.cindoc.csic.es/cybermetrics/articles/v5i1p2.html >.

Leydesdorff, Loet, and Peter Van den Besselaar (1994). *Evolutionary Economics and Chaos Theory: New Directions in Technology Studies*. London: Pinter.

Leydesdorff, Loet, and Nienke Oomes (1999) "Is the European Monetary System Converging to Integration?" *Social Science Information* 38(1): 57-86.
17

Leydesdorff, Loet, and Zeng Guoping (2001) "University-Industry-Government Relations in China: An emergent national system of innovations", *Industry and Higher Education* 15(3), 179-182.
Luhmann, Niklas (1984) *Soziale Systeme.Grundriss einer allgemeinen Theorie.* Frankfurt a.M.: Suhrkamp.
Lundvall, Bengt-Åke (1988) "Innovation as an interactive process: from user-producer interaction to the national system of innovation", in Dosi *et al.* (eds.), pp. 349-369.
McKelvey, Maureen D. (1996) *Evolutionary Innovations: The Business of Biotechnology.* Oxford: Oxford University Press.
McKelvey, Maureen D. (1997) "Emerging Environments in Biotechnology", in Henry Etzkowitz and Loet Leydesdorff (eds.), *Universities and the Global Knowledge Economy: A Triple Helix of University-Industry-Government Relations,* pp. ), 60-70.London: Cassell Academic.
Mowery, David C., and Nathan Rosenberg (1979) "The influence of market demand upon innovation: a critical review of some empirical studies", *Research Policy* 8: 102-153.
Nelson, Richard R. (ed.) (1982) *Government and Technical Progress: A cross-industry analysis.* New York: Pergamon.
Nelson, Richard R. (ed.) (1993) *National Innovation Systems: A comparative study.* Oxford and New York: Oxford University Press.
Nelson, Richard R. (1994) "Economic Growth via the Coevolution of Technology and Institutions", in Leydesdorff and Van den Besselaar (eds.), pp. 21-32.
Nelson, Richard R., and Sidney G. Winter (1982) *An Evolutionary Theory of Economic Change.* Cambridge, MA: Belknap Press.
Neurath, Otto, Rudolf Carnap, and Hans Hahn (1929) *Wissenschaftliche Weltauffassung — Der Wiener Kreis.* Vienna: Veröffentlichungen des Vereins Ernst Mach.
Noble, David (1979) *America by Design.* New York: Knopf.
Nowak, Michael J., and Charles E. Grantham (2000) "The virtual incubator: managing human capital in the software industry", *Research Policy* 29(2): 125-134.
Pavitt, Keith (1984) "Sectoral patterns of technical change: towards a theory and a taxonomy", *Research Policy* 13: 343-73.
Richta, Radovan *et al.* (1968) *Civilizace na rezcesti.* Prague [Richta-Report, *Politische Ökonomie des 20. Jahrhunderts: Die Auswirkungen der technisch-wissenschaftlichen Revolution auf die Produktionsverhältnisse.* Frankfurt a.M.: Makol, 1971].
Rip, Arie and Barend Van der Meulen (1996) "The Post-modern Research System," *Science and Public Policy* 23(6): 343-352.
Ronge, Volker (1979) *Bankpolitik im Spätkapitalismus. Starnberger Studien 3.* Frankfurt a.M.: Suhrkamp.
Scharnhorst, Andrea (1998) "Citation - Networks, Science Landscapes, and Evolutionary Strategies," *Scientometrics* 43: 95-139.
Schumpeter, Joseph (1939/1964) *Business Cycles: A Theoretical, Historical and Statistical Analysis of Capitalist Process.* New York: McGraw-Hill.
Simon, Herbert A. (1969) *The Sciences of the Artificial.* Cambridge: MA/ London: MIT Press.
Skolnikoff, Eugene B. (1993) *The Elusive Transformation.* Princeton, NJ: Princeton University Press.
Van den Belt, Henk, and Arie Rip (1987) "The Nelson-Winter-Dosi model and synthetic dye chemistry", in Bijker *et al.* (eds.), pp. 135-58.
Van den Daele, Wolfgang, Wolfgang Krohn, and Peter Weingart (eds.) (1979) *Geplante Forschung.* Frankfurt a.M.: Suhrkamp.
Weber, Max (1922/1972) *Wirtschaft und Gesellschaft.* Tübingen: Mohr.
Whitley, Richard D. (1984) *The Intellectual and Social Organization of the Sciences.* Oxford: Oxford University Press.

Woolgar, Steve (1988) *Science. The very idea.* London and New York: Tavistock Publications.

Wouters, Paul , Loet Leydesdorff, and Jan Annerstedt (1999) *The European Guide to Science, Technology and Innovation Studies.* Brussels: European Commission, DG XII; at http://www.chem.uva.nl/sts/guide/index.html.

Yamauchi, Ichizo (1986) "Long Range Strategic Planning in Japanese R&D", in Christopher Freeman (ed.), *Design, Innovation and Long Cycles in Economic Development,* pp. 169-185.London: Pinter.

Ziman, John (1994) *Prometheus Bound: Science in a Dynamic Steady State.* Cambridge: Cambridge University Press.